\documentclass[reprint, superscriptaddress, amsmath,amssymb, aps, prresearch, longbibliography, floatfix]{revtex4-2}
\usepackage{amsmath}
\usepackage{graphicx}
\usepackage{ulem}
\usepackage{bm}
\usepackage{natbib}
\usepackage{dsfont}
\usepackage{tikz}
\usepackage{epsfig}
\usepackage{feynmf}
\usepackage{blindtext, rotating}
\usepackage{mathtools}
\usepackage{dsfont}
\usepackage{subcaption}
\usepackage{physics}
\usepackage{amsfonts}
\usepackage{xcolor}
\usepackage{ragged2e}
\usepackage{siunitx}
\usepackage{comment}
\usepackage{soul}
\usepackage{lipsum}
\usepackage{hyperref} 
\hypersetup{breaklinks=true, colorlinks=true, citecolor=blue, linkcolor=cyan, urlcolor=blue,filecolor=blue}

\usepackage{xcolor} 

\DeclareCaptionJustification{justified}{\justifying}

\DeclareSIUnit{\rad}{rad}

\definecolor{bright_blue}{HTML}{85C1E9}
\definecolor{middle_blue}{HTML}{2E86C1}
\definecolor{dark_blue}{HTML}{1B4F72}

\begin{document}

\title{All-optical Saddle Trap}

\author{Daniel Tandeitnik}
\email{tandeitnik@gmail.com}
\affiliation{Department of Physics, Pontifical Catholic University of Rio de Janeiro, Rio de Janeiro 22451-900, Brazil}

\author{Oscar Kremer}
\affiliation{Department of Electrical Engineering, Pontifical Catholic University of Rio de Janeiro, 22451-900 Rio de Janeiro, RJ, Brazil}

\author{Felipe Almeida}
\affiliation{Department of Physics and Astronomy, University College London, Gower Street, WC1E 6BT London, UK}

\author{Joanna A. Zielińska}
\affiliation{Tecnologico de Monterrey, Escuela de Ingeniería y Ciencias, Monterrey, Nuevo León, México}

\author{Antonio Zelaquett Khoury}
\affiliation{Instituto de Física, Universidade Federal Fluminense, Niterói, Rio de Janeiro 24210-346, Brazil}

\author{Thiago Guerreiro}
\email{barbosa@puc-rio.br}
\affiliation{Department of Physics, Pontifical Catholic University of Rio de Janeiro, Rio de Janeiro 22451-900, Brazil}

\begin{abstract}
The superposition of frequency-shifted Laguerre-Gauss modes can produce a rotating saddle-like intensity profile. When spinning fast enough, the optical forces produced by this structured light saddle generate a dynamically stable equilibrium point capable of trapping nanoparticles in a high vacuum, akin to a Paul trap but with its unique characteristics. We analyze the stability conditions and center-of-mass motion, dynamics and cooling of a nanoparticle levitated in the optical saddle trap. We expect the optical saddle to find applications in levitated optomechanics experiments requiring fast parametric modulation and inverted squeezing potential landscapes.
\end{abstract}

\maketitle

\section{Introduction}



    

Trapping charged and neutral particles in electromagnetic fields has been one of the major tools in atomic, molecular, and nano-physics. While Earnshaw's theorem prevents the stable trapping of charged particles using electrostatic forces, stable equilibrium can be achieved dynamically by time-varying potentials \cite{kapitza1965dynamical}. These dynamic traps not only lie at the heart of trapped ion physics \cite{paul1990electromagnetic} but can also be used to trap mesoscopic objects such as charged dielectric nanoparticles \cite{bykov2019direct}. Neutral particles, on the other hand, can be trapped by purely electrodynamic fields, e.g. laser light, as first proposed by Ashkin \cite{ashkin1997optical}. Optical tweezers can operate in different regimes depending on the relative size of the particle compared to the laser wavelength, enabling the trapping of single atoms all the way to nano and microparticles \cite{gieseler2021optical}. Moreover, levitation in vacuum tweezers offers a promising platform for fundamental quantum physics and sensing experiments \cite{millen2020optomechanics, gonzalez2021levitodynamics}. Recently, hybrid optical-electrical traps combining both optical tweezers and dynamic electric traps have been designed and proposed \cite{bykov2022hybrid} for fundamental experiments seeking to delocalize the quantum state of a levitated mesoscopic object \cite{bonvin2024state, muffato2024generation}. 

\begin{figure}
    \includegraphics{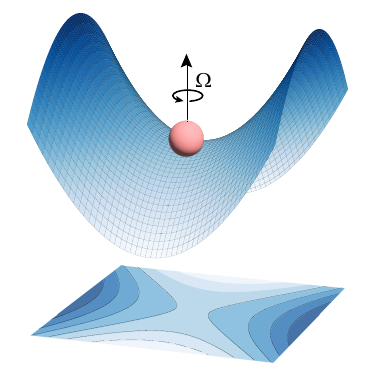}
    \caption{A saddle beam goes round and round, making a nanoparticle dynamically bound.}
    \label{fig:prettySaddle}
\end{figure}

Here we investigate whether the idea of a dynamic equilibrium trap can be implemented in an all-optical setup. Considering structured light both in space and in time, we propose a novel method for optical trapping of nano-sized dielectric particles in the dipole regime. By superposing frequency-shifted Laguerre-Gauss modes with a Gaussian beam, we can engineer a rotating saddle optical potential capable of holding a Silica nanoparticle in a high-vacuum environment, as illustrated in Figure \ref{fig:prettySaddle}. Provided the frequency shifts of the superposition components exceed a critical value, this optical saddle rotates at a sufficiently fast rate to develop a dynamic stable equilibrium point at the center of the beam. This structured light trap mimics a Paul trap but is made entirely with light and with a unique motion pattern and dynamics.

The all-optical saddle trap offers interesting possibilities for levitated optomechanics experiments, both classically and in future quantum experiments. A particle trapped in the rotating saddle can be cooled to a complex motional state using optimal control methods and electric feedback \cite{kremer2024all}. Once cooled, the trap's rotation frequency can be tuned from several hundreds of MHz to zero, causing an effective inverted potential upon the initially localized particle. This not only offers novel methods for state expansion \cite{bonvin2023hybrid, muffato2024generation} and interference protocols \cite{neumeier2024fast, weiss2021large}, but also the possibility of fast-switching between stable and unstable potential landscapes. 


The all-optical saddle trap offers exciting possibilities for both classical and future quantum levitated optomechanics experiments. Similar to traditional optical tweezers, particles can be cooled to their mechanical ground state using optimal control methods and electric feedback \cite{kremer2024all}. However, a key advantage of the saddle trap is its capability for rapid and precise switching between co-aligned harmonic and inverted potentials (or between different harmonic potential frequencies). This feature is essential for coherent state expansion \cite{bonvin2023hybrid, rossi2024quantumdelocalizationlevitatednanoparticle, weiss2021large} and interference protocols \cite{neumeier2024fast}. Unlike conventional optical tweezers, the saddle trap enables versatile potential switching through a single optical phase control, eliminating the need to vary the trapping beam pointing and power, which may lead to experimental artifacts.

This paper is organized as follows. In Sec. \ref{sec:theory}, we describe the optical saddle beam as well as the associated dynamics of a levitated nanoparticle. We derive and discuss the conditions required for achieving dynamic equilibrium and compute the power spectrum of the particle's motion in the laboratory, i.e., non-rotating frame of reference. Next, in Sec. \ref{sec:experimental}, we propose an experimental setup to implement the optical saddle trap and show that feedback cooling of the particle motion is possible by applying optimal control methods. We conclude with a discussion of the results.

\section{Theory}\label{sec:theory}

\subsection{The optical saddle beam}

There are different ways of producing a saddle intensity pattern around the focus of an optical beam. As a simple solution, we choose the superposition of three Laguerre-Gauss (LG) beams as follows:
\begin{align}\label{eq:saddleSuperposition}
    E_{\mathrm{s}} =  \sqrt{\frac{16P}{17c\epsilon_0}}\left(E^{LG}_{0,0}  + \frac{3}{4}e^{-i2\theta}E^{LG}_{0,2}   + \frac{3}{4}e^{i2\theta}E^{LG}_{0,-2}\right),
\end{align}
\noindent where $P$ is the laser power, $c$ is the speed of light in vacuum, $\epsilon_0$ is the electrical permittivity of the vacuum, and $\theta$ is a parameter that controls the orientation of the saddle on the transverse plane. Considering the optical-axis as the $z$-axis, the normalized LG mode $E^{LG}_{p,l}$ is defined in cylindrical coordinates as
\begin{align}
\begin{split}
    &E^{LG}_{p,l}(r,\phi,z) = \sqrt{\frac{2p!}{\pi(p+\vert l\vert)!}}\frac{1}{w(z)}\left(\frac{\sqrt{2}r}{w(z)}\right)^{\vert l\vert}  \\
    &\times L^{\vert l\vert}_p\left(\frac{2r^2}{w^2(z)}\right)\exp\left( \frac{-r^2}{w^2(z)} -ik\frac{r^2}{2R(z)} + il\phi+i\psi(z)\right),
\end{split}
\label{eq:LG_modes}
\end{align}
\noindent where  $w(z)$, $z_R$, $R(z)$, and $\psi(z)$ are the beam waist radius, Rayleigh range, wavefront radius and Gouy phase, respectively given by
\begin{subequations}
    \begin{align}
    &w(z) = w_0\sqrt{1+(z/z_R)^2}, \\
    &z_R = \pi w_0^2/\lambda_0, \\
    &R(z) = z(1+(z_R/z)^2), \\
    &\psi(z) = (\vert l\vert + 2p +1)\arctan(z/z_R),
\end{align}
\end{subequations}
\noindent with $w_0$ denoting the width of the Gaussian $L^{0}_0$ mode at the focus, $\lambda_0$ the beam's wavelength and $L^{\vert l\vert}_p$ are the generalized Laguerre polynomials. For the modes present in Eq. \eqref{eq:saddleSuperposition}, we have $L^{0}_0(x) = L^{\vert 2\vert}_0(x) = 1$.

\begin{figure}
    \includegraphics{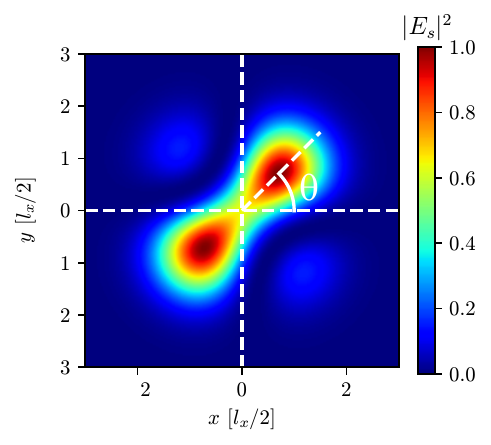}
    \caption{Transverse absolute square of the field at the focus. The values are normalized by the maximum value of $\vert E_s\vert^2$. The beam presents a saddle-like profile around the origin, and the parameter $\theta$ defines its orientation. Both axes are normalized by the distance between the origin and the bright side peaks.}
    \label{fig:intensitySaddle}
\end{figure}

Throughout this work, we assume a linearly polarized beam. Note that the saddle beam \eqref{eq:saddleSuperposition} and the LG modes \eqref{eq:LG_modes} are normalized so that the integral of the intensity $I = (c\epsilon_{0}/2)\vert E_{s} \vert^{2}$ over the transverse plane gives the total power $P$.

Evaluation of the absolute square of the field results in
\begin{align}\label{eq:intensitySaddle}
\begin{split}
     \vert E_{s}(r,\phi,z)\vert^2 &= \frac{16P}{17c\epsilon_0\pi w(z)^2}\exp\left(\frac{-2r^2}{w(z)^2}\right)\\&\times\Bigg[1+\frac{3\sqrt{2}r^2}{w(z)^2}\cos{(2\phi+2\theta)}\cos(2\chi(z))\\ &+ \frac{9r^4}{2w(z)^4}\cos^2(2\phi+2\theta)\Bigg],
\end{split}
\end{align}
\noindent where $\chi(z) = \arctan(z/z_R)$. Figure \ref{fig:intensitySaddle} shows the transverse absolute square of the field at the focus. 
The obtained intensity distribution corresponds to a saddle-like profile, whose orientation in the $xy$ plane depends on the relative phase $\theta$ between the Laguerre-Gauss $l=\pm 2$ and the Gaussian components of the beam [see Eq.~\eqref{eq:saddleSuperposition}]. The relationship between the orientation of the saddle pattern and the parameter $\theta$ is evident in Eq.~\eqref{eq:intensitySaddle},  where the azimuthal angle $\phi$ appears only as the sum $\phi+\theta$. Thus, as we sweep the phase $\theta$, the saddle rotates. 

We can express the intensity distribution from Eq.~\eqref{eq:intensitySaddle} in terms of  Cartesian coordinates rotating with the saddle, i.e. $x'=x\cos\theta+y\sin\theta$ and $y'=y\cos\theta-x\sin\theta$ where $x=r\cos\phi$ and $y=r\sin\phi$ are the stationary Cartesian coordinates, obtaining:
\begin{align}\label{eq:saddleIntensityRotFrame}
\begin{split}
     \vert E_{\mathrm{s}} (x',y',z)\vert^2 &= \frac{16P}{17c\epsilon_0\pi w(z)^2}\exp\left(\frac{-2(x'^2+y'^2)}{w(z)^2}\right)\\&\times\Bigg[1+\frac{6\sqrt{2}}{w(z)^2}\cos(2\chi(z))(x'^2-y'^2)\\ &+ \frac{18}{w(z)^4}(x'^2-y'^2)^2\Bigg].
\end{split}
\end{align}

\begin{figure}
    \includegraphics{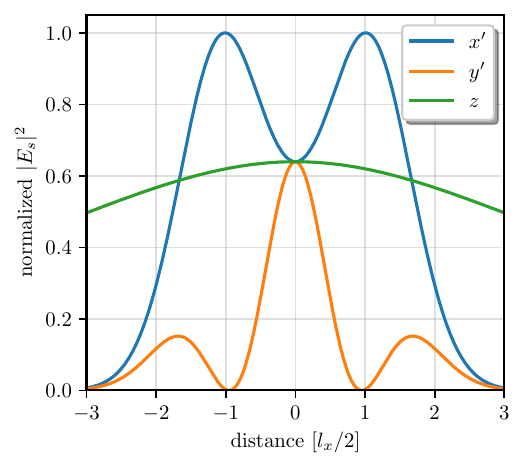}
    \caption{The absolute square of the saddle beam's electrical field along the Cartesian axes in the rotating reference frame. The intensity presents a saddle shape around the origin for the transversal plane.}
    \label{fig:intensityAlongAxis}
\end{figure}

Figure \ref{fig:intensityAlongAxis} shows the crosssections of $\vert E_{\mathrm{s}}\vert^2$ along the three axes of the aforementioned rotating reference frame. The separation between the peaks along the $x'$-axis $\ell_{x} $ equals 
\begin{eqnarray}
    \ell_{x} = \frac{w_0}{2}\sqrt{16-3\sqrt{2}}, 
\end{eqnarray}
and for the side peaks along the $y'$-axis $\ell_{y}$ equals
\begin{eqnarray}
    \ell_{y} = \frac{w_0}{2}\sqrt{16+3\sqrt{2}}.
\end{eqnarray}

\subsection{Dynamical model}

We now model the dynamics of a dielectric particle near the beam's focus. We consider a particle with a radius much smaller than the beam's wavelength, the dipole approximation. In this case, one considers the dielectric limit where the optical force exerted by the electric field onto the particle equals \cite{jones2015optical},
\begin{align}
    \label{eq:opticalForce}
    \mathbf{F}_{\rm{opt}} = \frac{\Re\{\alpha\}}{4}\nabla\vert E_s\vert^2 +\frac{\sigma_{\rm{ext}}}{c}\mathbf{S},
\end{align}
\noindent where $\alpha$ is the complex polarizability of the particle, $\sigma_{\rm{ext}}$ is the particle's extinction cross-section, and $\mathbf{S}$ is the time-averaged Poynting vector. We omit the spin-curl force since we are interested in uniformly linearly polarized beams. The first term in Eq. \eqref{eq:opticalForce} is commonly referred to as the gradient force, while the second is known as the scattering force. In the dipole approximation, the gradient force traps a particle near the focus of an optical tweezer. The scattering force, on the other hand, pushes the particle along the optical axis. To achieve stable trapping, the scattering must be smaller than the gradient force. For the saddle beam, the scattering force is mainly provided by the Gaussian amplitude $E^{LG}_{0,0}$ in Eq. \eqref{eq:saddleSuperposition} and displaces the equilibrium position of the particle along the optical axis. We take the scattering force into account throughout our simulations of the particle dynamics. 

From the plots of $\vert E_s\vert^2$ Figure~\ref{fig:intensityAlongAxis}, one sees that the saddle intensity in the transverse plane creates an unstable equilibrium point at the origin. Therefore, if a dielectric particle is placed at the focus and undergoes small perturbations, it will be pushed to one of the side peaks on the $x'$-axis. In analogy to electrical quadrupole and mechanical saddle traps which can confine particles in dynamical saddle-shaped potentials \cite{rueckner1995rotating,hasegawa2005stability,kirillov2016rotating}, we investigate the dynamics of a dielectric nanoparticle in the rotating optical saddle beam, i.e., by setting $\theta = \Omega t$, where $\Omega$ is the angular velocity of rotation. As we will see, provided $ \Omega $ is large enough, the unstable origin is turned into a dynamically stable equilibrium point.


The dynamics of a dielectric nanoparticle in the saddle becomes easier to analyze in a frame rotating about the $z$-axis with angular velocity $\Omega$ for which the intensity is given by Eq. \eqref{eq:saddleIntensityRotFrame}. The Taylor expansion of this intensity near the origin, retaining at most quadratic terms, yields
\begin{align}
\begin{split}
     &\vert E_{s}\vert^2\approx \frac{32P}{17c\epsilon_0\pi w_0^2} \\ &\times\left(1 + \frac{x^2}{w_0^2}(3\sqrt{2}-2) - \frac{y^2}{w_0^2}(3\sqrt{2}+2) - \frac{z^2}{z_R^2}\right)\label{eq:TaylorExpansionSaddle}.
\end{split}
\end{align}
At this point, we wish to derive a stability criterion for the conservative gradient force, and for this reason, we will neglect the scattering force for the moment. Note, however, that the effects of scattering upon the nanoparticle's dynamical equilibrium are discussed in Appendix \ref{sec:ScatterAppendix} and taken into account in the dynamical simulations shown in the next sections.

To linear order in the particle's displacement, the gradient force $\mathbf{F}_g$ given by the first term in Eq. \eqref{eq:opticalForce} reads
\begin{equation}
    \mathbf{F}_g = \frac{16P\Re\{\alpha\}}{17c\epsilon_0\pi w_0^2}
    \begin{bmatrix}
        x(3\sqrt{2}-2)/w_0^2 \\
        -y(3\sqrt{2}+2)/w_0^2 \\
        -z/z_R^2
    \end{bmatrix}.
\end{equation}

Since we describe the dynamics in a rotating reference frame, we must account for the centripetal and Coriolis forces that arise in such a frame. Newton's second law gives
\begin{equation}
    m\frac{d^2\mathbf{r}}{dt^2}\Bigg|_S = \sum_i\mathbf{F}_i(\mathbf{r},\dot{\mathbf{r}})\Bigg|_S,
\end{equation}
\noindent where the $S$ stands for the inertial frame of the lab. To write this equation in a rotating reference frame $S'$ that rotates about a vector $\mathbf{\Omega}$, one needs to rewrite the derivative of vectors as \cite{taylor2005classical}
\begin{equation}\label{eq:rotatingDerivative}
    \frac{d\mathbf{A}}{dt}\Bigg|_S =  \frac{d\mathbf{A}}{dt}\Bigg|_{S'} + \mathbf{\Omega}\times\mathbf{A}.
\end{equation}
\noindent Considering constant angular velocity, $d\mathbf{\Omega}/dt = 0$ we find
\begin{equation}\label{eq:newton2lawRotated}
    m\frac{d^2\mathbf{r}'}{dt^2}\Bigg|_{S'} = \sum_i\mathbf{F}_i(\mathbf{r}',\dot{\mathbf{r}}')\Bigg|_{S'} - 2m\mathbf{\Omega}\times\frac{d\mathbf{r}'}{dt}\Bigg|_{S'} - \mathbf{\Omega}\times(\mathbf{\Omega}\times\mathbf{r}').
\end{equation}
\noindent where $\mathbf{\Omega} = \Omega \hat{\mathbf{z}}$. Moreover, we consider a damping force in the lab frame of the form $-m\gamma\Dot{\mathbf{r}}$, where $m$ is the particle's mass and $\gamma$ is the drag coefficient. In the rotating reference frame this damping force becomes $-m\gamma\Dot{\mathbf{r}}' -m\gamma \Omega\times\mathbf{r}'$. 

Putting it all together, Eq. \eqref{eq:newton2lawRotated} defines a set of three differential coupled equations describing the motion of the particle,
\begin{subequations}\label{eq:UnderdampedEq}
    \begin{align}
        \Ddot{x}' &= (\Omega^2+\omega_x^2)x' - \gamma(\Dot{x}'-\Omega y') + 2\Omega\Dot{y}' + \frac{F_{x}}{m}\label{eq:dampedXeq},\\
        \Ddot{y}' &= (\Omega^2-\omega_y^2)y' - \gamma(\Dot{y}'+\Omega x') - 2\Omega\Dot{x}' + \frac{F_{y}}{m}\label{eq:dampedYeq}, \\
        \Ddot{z} &= -\omega_z^2 z - \gamma \Dot{z} +  \frac{F_{z}}{m}\label{eq:dampedZeq},
    \end{align}
\end{subequations}
where
\begin{equation}\label{eq:omegaDefinition}
    \begin{bmatrix}
        \omega_x^2 \\
        \omega_y^2 \\
        \omega_z^2
    \end{bmatrix}
    =
    \frac{16P\Re\{\alpha\}}{17c\epsilon_0\pi w_0^2m}
    \begin{bmatrix}
        (3\sqrt{2}-2)/w_0^2 \\
        (3\sqrt{2}+2)/w_0^2 \\
        1/z_R^2
    \end{bmatrix},
\end{equation}
\noindent and $F_{i} \ (i = x,y,z)$ represent additional force terms the particle might be subject to, such as for example thermal stochastic or feedback forces. For now, we consider $F_{i} = 0$ and derive a criterion for stable trapping in the saddle beam. 


\subsection{Stability analysis}\label{sec:dampedStability}

We now consider motion along the transverse plane. Instead of solving for $x'$ and $y'$ \footnote{Note that a general solution can be found by writing the equations as a set of four coupled first-order differential equations and employing diagonalization. The result, however, is extremely cumbersome and not physically illuminating.}
we seek a stability condition on the rotation frequency $\Omega$. Regarding the longitudinal $z$ direction in the presence of the scattering force, one can show that the particle follows a simple damped harmonic motion, see Appendix \ref{sec:ScatterAppendix}) for details.

Consider the ansatz $x' = x'_0e^{\lambda t}$ and $y' = y'_0e^{\lambda t}$. For confinement in the transverse plane, we require the values of $\lambda$ to have negative or null real part. Substituting  the ansatz in Eqs. \eqref{eq:dampedXeq}, \eqref{eq:dampedYeq} and casting the result in matrix form we find
\begin{equation}
    \begin{bmatrix}
        \Omega^2+\omega_x^2-\lambda^2-\gamma\lambda & \gamma\Omega+2\lambda\Omega\\
        -\gamma\Omega-2\Omega\lambda & \Omega^2-\omega_y^2-\lambda^2-\gamma\lambda
\end{bmatrix}
\begin{bmatrix}
        x'_0\\
        y'_0
\end{bmatrix} = \begin{bmatrix}
        0\\
        0
\end{bmatrix}.
\label{eq:stab_conditions}
\end{equation}
\noindent The existence of nontrivial solutions require the above matrix to be non-invertible (vanishing determinant), which gives the characteristic equation
\begin{align}
    \begin{split}
        \lambda ^4 &+ 2 \gamma  \lambda ^3 + \lambda ^2 \left(\gamma ^2+2 \Omega ^2+\omega_y^2-\omega_x^2\right) \\
        &+\lambda  \left(2 \gamma  \Omega ^2+\gamma  \omega_y^2-\gamma  \omega_x^2\right) \\
        &+\Omega ^4+\Omega ^2(\gamma ^2 +\omega_x^2-\omega_y^2)-\omega_x^2 \omega_y^2 = 0.
    \end{split}
\end{align}
\noindent with solutions given by
\begin{subequations}
    \begin{align}
        &\lambda_1^\pm \to \frac{1}{2} \left(-\gamma\pm\sqrt{2 \sqrt{a}+b} \right),\label{eq:firstRootsDamped}\\
        &\lambda_2^\pm \to \frac{1}{2} \left(-\gamma\pm\sqrt{\gamma ^2-2 \left(\sqrt{a}+c\right)}\right)\label{eq:secondRootsDamped},
    \end{align}
\end{subequations}
where
\begin{subequations}
\begin{align}
     a &= 8\Omega^2(- \gamma ^2/2+\omega_y^2-\omega_x^2)+(\omega_x^2+\omega_y^2)^2,\\
    b &= \gamma ^2-4 \Omega ^2+2 \omega_x^2-2 \omega_y^2, \\
    c &= 2 \Omega ^2-\omega_x^2+\omega_y^2.
\end{align}
\end{subequations}

The $ \lambda_{1}^{-} $ root in Eq. \eqref{eq:firstRootsDamped} is negative. On the other hand, $ \lambda_{1}^{+} $ will be negative when
\begin{equation}
    -\gamma+\sqrt{2 \sqrt{a}+b} < 0.
\end{equation}
\noindent Given $\omega_y > \omega_x$ (see Eq. \eqref{eq:omegaDefinition}), this condition implies
\begin{equation}
    \vert\Omega_c\vert > \sqrt{\frac{-\gamma^2+\omega_y^2-\omega_x^2+\sqrt{(\gamma ^2+\omega_x^2-\omega_y^2)^2+4\omega_x^2 \omega_y^2}}{2}}.
    \label{eq:stability_criteria}
\end{equation}
\noindent Therefore, the first two roots are negative provided $\Omega$ is sufficiently large and $\Omega_c$ defines the critical rotation frequency.

We now turn to the remaining roots; as before $ \lambda_{2}^{-} $ is negative, while $ \lambda_{2}^{+} $ will be negative when
\begin{equation}
    -\gamma+\sqrt{\gamma ^2-2 \left(\sqrt{a}+c\right)} < 0.
\end{equation}
\noindent As it turns out, this inequality is always satisfied for $ \gamma, \omega_{x}, \omega_{y} > 0$ and $\omega_y > \omega_x$. Hence, the stability criterion is given entirely by Eq. \eqref{eq:stability_criteria}. Considering the dynamics in the overdamped regime, the same criteria are obtained much more straightforwardly; see Appendix \ref{sec:OverdampedAppendix} for more details.

\subsection{Dynamics simulation}\label{sec:simulations}

\begin{figure*}[t]
    \includegraphics{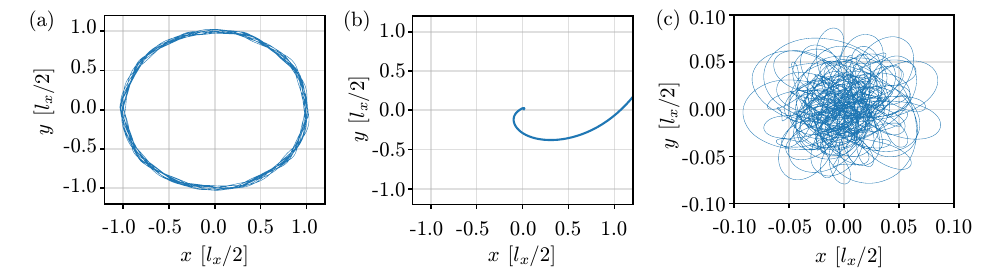}
    \caption{Representative trajectories of the three possible dynamics as viewed in the laboratory reference frame. Trajectory shows the last $\SI{0.4}{ms}$ transversal positions. a) $\Omega/\Omega_c = 0.1$.  The particle describes a closed trajectory following one of the rotating side peaks of the saddle. b) $\Omega/\Omega_c = 0.5 $. The particle escapes if the angular rotational velocity is not slow enough and well below the critical value for stability. c) $\Omega/\Omega_c = 1.0$. The particle displays dynamical stability and undergoes a complex trajectory about the saddle's center.}
    \label{fig:simulationCases}
\end{figure*}

We have studied the dynamics of a trapped particle in the saddle beam as well as the validity of the stability criteria by numerical simulation. In doing so, we have considered both the linearized model given by Eqs. \eqref{eq:UnderdampedEq} and the complete form of the potential given by the gradient of Eq. \eqref{eq:saddleIntensityRotFrame} for the fixed set of experimental parameters defined in Table \ref{tab:simulationParameters}. In all simulations, the particle initializes near the origin, where a quadratic potential landscape well approximates the trap. All simulations consider a stochastic external force $F_i = F_{\rm{th,i}}$ arising from collisions of the particle with surrounding gas molecules. This stochastic force obeys,
\begin{subequations}\label{eq:thermalCorrelation}
\begin{align}
    \langle F_{\rm{th,i}}(t)\rangle &= 0,\\
    \langle F_{\rm{th,i}}(t) F_{\rm{th,j}}(t+\tau)\rangle &= 2 m\gamma k_B T\delta(\tau)\delta_{ij},
\end{align}
\end{subequations}
\noindent where $k_B$ is the Boltzmann constant and $T$ is the surrounding gas temperature. We employed the Runge-Kutta $4^{\rm th}$-order algorithm for all simulations, where each run simulated the dynamics for a period of $\SI{10}{ms}$. We refer to Appendix \ref{sec:simulationsDetails} for more details.

\begin{table}[ht!]
\begin{tabular}{lcc}
\hline \hline
Parameter &\,\,  &Value \\ \hline
Particle composition &\,\, &SiO2\\
Particle radius &\,\, &$\SI{150}{nm}$  \\
Particle mass &\,\, &$\approx\SI{26}{fg}$  \\
Initial velocity ($\approx \sqrt{k_{B}T / m}$)&\,\,\,\,\,\,\,\,\,\,\,\,\,\,\,\,\,\,\,\,\,\,\,\,\,\,\,\,\,\,\,\,\,\,\,\,\,\,\,\, &$\approx\SI{12.5}{\mu m/s}$  \\
Laser wavelength &\,\, &$\SI{1550}{nm}$  \\
Tweezer power &\,\, &$\SI{500}{mW}$  \\
Trapping NA &\,\, &$0.6$  \\
Pressure &\,\, &$\SI{10}{mbar}$ \\
Temperature &\,\, &$\SI{293}{K}$ \\
\hline \hline 
\end{tabular}
\caption{Experimental parameters considered in the dynamical simulations.}
\label{tab:simulationParameters}
\end{table}

\begin{figure}
    \includegraphics{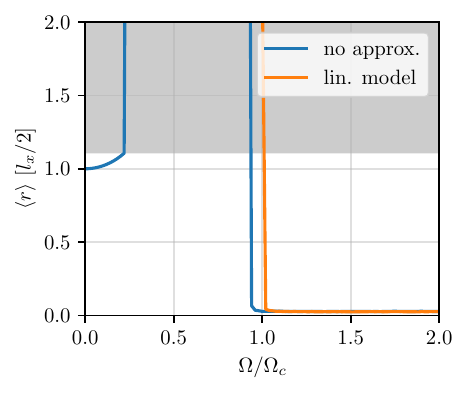}
    \caption{Stability criteria, both for the linearized model (orange line) and full optical potential (blue line).  
    The mean value of the last $\SI{1}{ms}$ of the particle's radial position is plotted as a function of the ratio between the saddle's rotational and critical angular velocities $\Omega/\Omega_c$. The shaded region indicates the particle is no longer trapped.}
    \label{fig:fullTaylorComp}
\end{figure}

To evaluate the different regimes of motion of a particle in the rotating saddle, we have considered different ratios of the trap's rotating frequency $\Omega$ to the critical angular velocity $\Omega_c$ given by Eq. \eqref{eq:stability_criteria}. Figure \ref{fig:simulationCases} qualitatively displays three regimes of motion: a) If $\Omega$ is slow enough ($\Omega/\Omega_c \lesssim 0.25$), the particle is trapped by one of the rotating side-peaks, following an approximately circular motion. As the rotation is increased, the particle lags behind the peak due to the drag force provided by the surrounding gas; b) increasing the rotation but keeping it well below the critical angular velocity, the centrifugal force as felt by the particle in the rotating frame becomes so strong that the particle is expelled away from the beam. Going higher in angular velocity c), at $\Omega/\Omega_c \gtrsim 0.92$, the particle is confined in the center and undergoes a complex motion around this dynamically generated equilibrium point. We note that for the parameters in Table~\ref{tab:simulationParameters} we have $\Omega_c/2\pi \approx \SI{280}{kHz}$. All trajectories are viewed in the laboratory reference frame.
\begin{figure*}[t!]
    \includegraphics{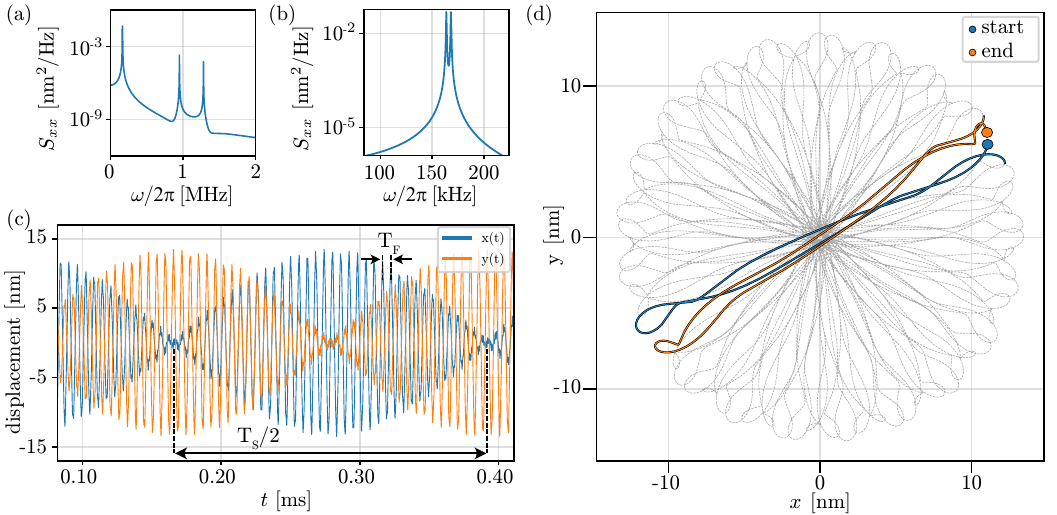}
    \caption{a) PSD for the trapped particle's $x $ motion as calculated by linearizing the saddle beam intensity. b) A zoom of the $ x $ motion PSD shows that at low frequencies, two peaks arise from the motion. c) Trajectories of the trapped particle along the $ x $ and $ y $ coordinates, showing the fast $ T_{F}$ and slow $ T_{S}$ periods of oscillation that originate the two low-frequency peaks in the PSD. d) The trajectory of the particle in the $x-y$ plane shows both slow and fast oscillations, as well as higher frequency micro-motions which create additional peaks in the power spectral density. The trajectory consists of 37 fast oscillations, starting at the blue point and ending at the orange point. A complete fast oscillation is represented by the blue and orange curves, while the intermediate oscillations are shown by the gray dotted curve. During the fast oscillations, the orbit gradually twists around the z-axis, resulting in the slow oscillation.}
    \label{fig:trajectory_analysis_panel}
\end{figure*}
Figure \ref{fig:fullTaylorComp} shows an alternative visualization of the different regimes of motion. Here, the particle's last radial position $\langle r \rangle = \langle\sqrt{x^2+y^2}\rangle$ output by the simulation is averaged over 20 runs and plotted against the $\Omega/\Omega_c$ ratio. 

We see that for $\Omega/\Omega_c > 1 $, the particle remains trapped near the center of the beam. On the other hand, for $\Omega/\Omega_c < 1 $, the value of $\langle r \rangle $ blows up, indicating that the particle is unbounded (shaded region). For sufficiently small values of $\Omega$, the particle is bound in one of the lobes of the saddle beam, remaining at a position of the intensity maxima given by $ l_{x}/2 $.

We now turn to analyze the transversal trajectory dynamics of the trapped particle. Commonly employed in levitated optomechanics experiments as a characterization method \cite{bykov2022hybrid}, the power spectrum density (PSD) of the motion of a particle trapped in the rotating saddle can be analytically computed by linearizing the Eqs. of motion. Appendix \ref{sec:PSD} contains the calculation details for the rotating and laboratory reference frames. Figure \ref{fig:trajectory_analysis_panel}a depicts an example of the analytical prediction of the particle's $x$ position PSD in the laboratory frame for the parameters contained in Table \ref{tab:simulationParameters} and $\Omega/\Omega_c = 2$. The PSD exhibits four peaks, two lying very close together, as seen from the zoom over the $100$ - $\SI{200}{kHz}$ range in Figure \ref{fig:trajectory_analysis_panel}b. These neighbor peaks arise from a beating pattern in the motion along the  $x$ and $ y $ axes, as seen in Figure \ref{fig:trajectory_analysis_panel}(c), where we see a simulated trajectory in the absence of the drag force. The inverse of half the sum (difference) of the neighbor peak frequencies defines a fast (slow) period $T_F$ ($T_S$) of oscillation. Together with the fact that the $x$ and $y$ motions are phase-locked, the result is a Lissajous-like curve in the transversal direction, as depicted in Figure \ref{fig:trajectory_analysis_panel}(d). The fast oscillation gives rise to a squashed orbit around the origin, which slowly rotates around the $z$-axis. After half a slow period, the trajectory is close to where it started, where the orbit will be closed if $T_S/(T_F/2)$ is an integer number. Finally, the remaining high-frequency peaks of the PSD shown in Figure \ref{fig:trajectory_analysis_panel}(a) are responsible for the micro-movements of the trajectory seen in \ref{fig:trajectory_analysis_panel}(d).




\section{Experimental proposal}\label{sec:experimental}

We now discuss an experimental scheme for generating the rotating saddle beam as well as the possibility of feedback cooling the center-of-mass motion of a trapped nanoparticle.

\subsection{Saddle-beam generation}


To generate the superposition in Eq.\eqref{eq:saddleSuperposition} rotating with angular velocity in the MHz range, we propose the combined use of three acoustic optical modulators (AOM) and a variable spiral plate (VSP) -- a liquid crystal element capable of transforming an incident circularly polarized Gaussian beam into Laguerre-Gaussian modes with $l = \pm 2$ depending on the input's handedness \cite{rubano2019q}.

\begin{figure}
    \includegraphics{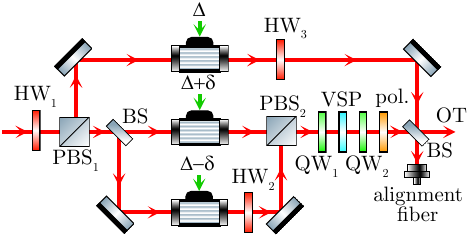}
    \caption{Experimental proposed setup to generate a fast-rotating saddle beam. Three AOMs are employed to give the correct relative phase to the components of the saddle beam superposition. Gaussian beams with opposite circular polarization generate the required $ l =\pm 2$ LG modes using a VSP.}
    \label{fig:experimentalSetup}
\end{figure}

Figure \ref{fig:experimentalSetup} illustrates the proposed optical setup. The half-wave plate $\rm{HW}_1$, in conjunction with the polarizing beam splitter $\rm{PBS_1}$, divides an initial linearly polarized laser beam into two paths. The upper path receives a frequency shift of $\Delta$, and $\rm{HW}_3$ corrects the polarization for proper interference at the optical tweezer (OT) site. The lower path is divided into two beams that receive distinct frequency shifts of $\Delta \pm \delta$ and are recombined with $\rm{HW}_2$ and $\rm{PBS}_2$ into a single spatial mode. Since each component has orthogonal linear polarizations, they have opposite handedness after passing through the quarter-wave plate $\rm{QW}_1$. Therefore, the VSP transforms one of the beams into a $l = 2$ and the other into a $l = -2$ LG mode. Finally, $\rm{QW}_2$ reverts the circular polarization to linear, and the polarizer projects into a common mode to interfere with the remaining Gaussian beam. The distinct detunings received by the superposition components effectively generate a time-dependent phase, causing the overall superposition to rotate. Tuning the frequency shifts of each AOM by $ \delta $ around a central shift $ \Delta $ allows for fine control of the saddle's rotation speed.
Coupling all three beams into a single-mode alignment fiber before inserting the VSP ensures the spatial mode matching of the rotating saddle's components. 

Setting the transmission:reflection ratio of $\rm{HW}_1$ to approximately 70:30, the saddle beam can be reproduced with the proper coefficients. Considering the AOM's diffraction efficiency as $80\%$ and the initial beam's power as $\SI{1}{W}$, roughly $\SI{260}{mW}$ of power should arrive at the OT. We note that other schemes can also be considered, such as using a single AOM with a double pass. However, in practice, most schemes have severely low power efficiencies at the output. Schemes using one or two SLMs to generate the LG beams were also disfavored in favor of the VSP since the latter is a low-loss optical element. 

Finally, commercially available AOMs can shift the frequency of an incoming laser beam by tens of MHz around a central frequency $\Delta$. Thus, by choosing a shift of $\delta = \SI{10}{MHz}$, the saddle beam would rotate at sufficient angular velocity to have dynamical stability.

\subsection{Feedback cooling}

For most applications in high-vacuum levitated optomechanics, the ability to control the motion of the trapped particle is required. For instance, optimal control theory can successfully achieve the ground state of the particle's longitudinal center-of-mass (CoM) motion \cite{magrini2021real, tebbenjohanns2021quantum}. We thus ask the question of whether optimal control theory methods can be applied to cool the motion of a levitated nanoparticle in the rotating saddle.

The issue of minimizing the energy of linear systems, as the one presented in Eq. \eqref{eq:UnderdampedEq}, can be effectively addressed using the optimal control policy known as Linear Quadratic Gaussian (LQG) controller. The dynamics of a trapped particle in the rotating reference frame can be approximated by a set of linear differential equations, and the implementation of the feedback loop is highly reliant on the detection of the particle's position. In our case, the particle's positions are detected in the laboratory frame by means of homodyne detection of the trap's scattered light.
One can express the detection signals by
\begin{equation}\label{eq:measurement-equation}
    \mathbf{y}(t)=\mathbf{C}\mathbf{x}(t)+\mathbf{m}(t),
\end{equation}
\noindent where we employ the state-space representation, in which $\mathbf{x}(t)$ is the state vector for the laboratory reference frame, defined as
\begin{equation}
    \mathbf{x}(t)=\begin{bmatrix}
    x(t) & y(t) & z(t) & \dot{x}(t) & \dot{y}(t) & \dot{z}(t)
\end{bmatrix}^T,
\end{equation}
\noindent and $\mathbf{y}(t)$ is the measurement vector. Moreover, $\mathbf{C}$ is a $3\times 6$ matrix defined by the calibration factors of the detectors and $\mathbf{m}(t)$ is the measurement noise vector, a $3\times 1$ vector composed by independent white Gaussian noise processes satisfying $\langle \mathbf{m}(t)\rangle = \mathbf{0}_{3\times 1}$ and $ \langle\mathbf{m}(t)\mathbf{m}(t)^T\rangle=\mathbf{M}$, where $\mathbf{M}$ is a $3\times 3$ diagonal matrix.

\begin{figure*}[!htb]
    \centering
    \includegraphics[]{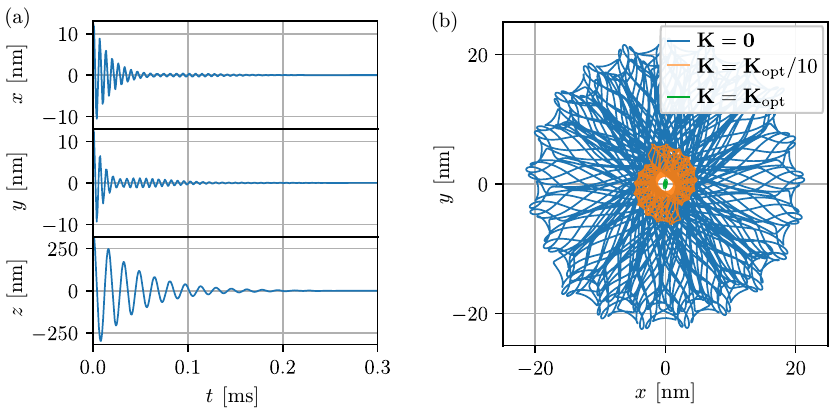}
    \caption{Optimal control for an optical saddle trap. a) Time traces for $x$, $y$ and $z$ coordinates of a particle levitated by an optical saddle trap and controlled by the LQG. b) Motion of a trapped particle along the transverse plane for different feedback gains. Numerical simulations were carried considering SiO2 nanoparticles with a radius of $\SI{150}{nm}$, pressure of $\SI[parse-numbers=false]{10^{-5}}{mbar}$, gas temperature of $\SI{293}{K}$ and detection efficiencies of $1.0\%$, $1.0\%$, and $0.1\%$ for $x$, $y$, and $z$, respectively.}
    \label{fig:panel_control}
\end{figure*}

Using Eq. \eqref{eq:cordinateTransformation}, one finds the relation between $\mathbf{x}$ and the state vector in the rotating frame $\mathbf{x}^\prime$,

\begin{equation}\label{eq:relation-states-frames}
    \mathbf{x}(t)=\mathbf{T}(\theta)\mathbf{x}^\prime(t),
\end{equation}
where $\mathbf{T}(\theta)$ is given by
\begin{equation}
    \mathbf{T}(\theta)=\begin{bmatrix}
        \mathbf{R}_z(\theta) & \mathbf{0}_{3\times 3} \\
        \dot{\mathbf{R}}_z(\theta) & \mathbf{R}_z(\theta)
    \end{bmatrix},
\end{equation}
with,
\begin{equation}
    \mathbf{R}_z(\theta) = \begin{bmatrix}
    \cos\theta & -\sin\theta & 0 \\
    \sin\theta & \cos\theta & 0 \\
    0 & 0 & 1
\end{bmatrix}.
\end{equation}
\noindent From Eq. \eqref{eq:relation-states-frames}, we can write an expression analogous to the one in Eq. \eqref{eq:measurement-equation} for the rotating reference frame, yielding 
\begin{equation}\label{eq:measurement-rotating}
    \mathbf{y}^\prime(t) = \mathbf{C}^\prime\mathbf{x}^\prime(t) +\mathbf{m}^\prime(t),
\end{equation}
where 
\begin{equation}
\mathbf{y}^\prime(t)=\mathbf{R}_z^T(\theta)\mathbf{y}(t) \, \textrm{ ,} \, \mathbf{C}^\prime=\mathbf{R}_z^T(\theta)\mathbf{C}\mathbf{T}(\theta).
\end{equation}
\noindent The noise vector $\mathbf{m}^\prime$ has zero mean and satisfies 
\begin{equation}
    \langle \mathbf{m}^\prime(t)\mathbf{m}^{\prime}(t)^T\rangle=\mathbf{R}_z^T(\theta)\mathbf{M}\mathbf{R}_z(\theta).
\end{equation}


Additionally, the state-space representation also allows the rewriting of the dynamics presented in Eq. \eqref{eq:UnderdampedEq}, yielding
\begin{equation}\label{eq:dynamics-statespace-rotating}
\dot{\mathbf{x}}^\prime(t)=\mathbf{A}^\prime\mathbf{x}^\prime(t) + \mathbf{B}^\prime\mathbf{u}^\prime(t)+\mathbf{w}^\prime(t),
\end{equation}
with $\mathbf{w}^\prime(t)$ taking into account the stochastic forces acting upon the particle, $\mathbf{B}^\prime \mathbf{u}(t)$ being related to the feedback forces exerted by the controller and $\mathbf{A}^\prime$ the state matrix, defined as 
\begin{equation}
    \mathbf{A}^\prime=
    \begin{bmatrix}
        0 & 0 & 0 & 1 & 0 & 0 \\
        0 & 0 & 0 & 0 & 1 & 0 \\
        0 & 0 & 0 & 0 & 0 & 1 \\
        \Omega^2+\omega_x^2 & \gamma\Omega & 0 & -\gamma & 2\Omega & 0 \\
        -\gamma\Omega & \Omega^2-\omega_y^2 & 0 & -2\Omega & -\gamma & 0 \\
        0 & 0 & -\omega_z^2 & 0 & 0 & -\gamma \\
    \end{bmatrix}.
\end{equation}

Equations \eqref{eq:measurement-rotating} and \eqref{eq:dynamics-statespace-rotating} form a fundamental pair in linear control theory, from where one extracts fundamental information about the system's dynamics, actuators, and detection. Such information must be properly characterized to apply LQG correctly. The optimal control law $\mathbf{u}^{\prime\ast}$  is  then
\begin{equation}
    \mathbf{u}^{\prime\ast}=-\mathbf{K}_{\rm{opt}}\hat{\mathbf{x}}^\prime(t),
\end{equation}
where $\mathbf{K}_{\rm{opt}}$ is the optimal controller's gain matrix and $\hat{\mathbf{x}}^\prime$ is an estimation of the state vector returned by the Kalman-Bucy filtering method \cite{aastrom2012introduction}. This filtering approach has been shown to be equivalent of the quantum filter for Gaussian systems \cite{belavkin1999measurement} and has enable longitudinal ground state cooling of a levitated nanoparticle \cite{magrini2021real}. By extending its application to linear dynamics in non-inertial frames, we enable its use in controlling the nanoparticle's motion in the saddle trap, while preserving its role as a classical analogue to quantum dynamics. Physically, the control law can be implemented using electrodes near the trap center \cite{kremer2024all}. Note that the computation of $\mathbf{K}_{\rm{opt}}$ and $\hat{\mathbf{x}}^\prime(t)$ depends on knowledge of $\mathbf{A}^\prime$, $\mathbf{B}^\prime$, $\mathbf{C}^\prime$, $\mathbf{w}^\prime(t)$ and $\mathbf{m}^\prime(t)$. For more details regarding the estimations and optimal gain calculations, we refer to \cite{kirk2004optimal}. 

We performed numerical simulations to examine the effectiveness of the LQG in controlling a particle confined by the saddle beam, assuming the same particle parameters as shown in Table \ref{tab:simulationParameters}. Figure \ref{fig:panel_control} a) presents the simulation results for the three directions in the lab fame. We observe a significant reduction in the amplitude of the particle's CoM amplitude for all three directions of motion as time evolves. The motion reaches a stationary state near $\SI{0.2}{ms}$ after starting the application of the control law. Figure  \ref{fig:panel_control}b) shows the controller's effect on the transversal plane motion; we see the motion amplitude increase as we apply smaller feedback gains. Finally, note that we carried out the simulations considering the complete form of the potential, which is highly nonlinear. Even though the LQG formalism utilizes linearized dynamics, it is still effectively capable of appreciably damping the particle's motion.


\section{Conclusion}

In conclusion, we proposed a rotating saddle-like structured beam for optical levitation experiments. When rotating above a critical velocity, the saddle beam acquires a dynamical equilibrium point capable of levitating dielectric nanoparticles in high-vacuum. The power spectrum of the center-of-mass motion presents unique features associated with the non-harmonic nature of the trap. We have proposed a method to generate the rotating saddle with a controllable rotation velocity and have shown that feedback control can be used to cool the motion of the particle. We expect the saddle beam to find applications in fundamental quantum physics experiments using levitated nano-objects, where inverted and nonlinear potentials can be used to rapidly expand and interfere with an initially localized wavepacket \cite{weiss2021large, neumeier2024fast}.
\\
\section*{Funding and Competing Interests}

This work was supported by the Coordena\c{c}\~ao de Aperfei\c{c}oamento de Pessoal de N\'ivel Superior - Brasil (CAPES) - Finance Code 001, Conselho Nacional de Desenvolvimento Cient\'ifico e Tecnol\'ogico (CNPq scholarship 140197/2022-2), Funda\c{c}\~ao de Amparo \`a Pesquisa do Estado do Rio de Janeiro (FAPERJ Scholarship No. E-26/200.252/2023, E-26/202.330/2024 and E-26/202.762/2024), Funda\c{c}\~ao de Amparo \`a Pesquisa do Estado de São Paulo (FAPESP process No. 2021/06736-5 and 2021/06823-5), the Serrapilheira Institute
(grant No. Serra – 2211-42299) and StoneLab. We acknowledge support from EPSRC International Quantum Technology Network LeviNet EP/W02683X/1.

All authors certify that they have no affiliations with or involvement in any organization or entity with any financial or non-financial interest in the subject matter or materials discussed in this manuscript.

\bibliography{main}

\newpage

\appendix

\section{The scattering force}\label{sec:ScatterAppendix}

To determine the stability condition for the $z$ direction the scattering force needs to be analysed. Stable equilibrium occurs if there is a point where the scattering and gradient forces are opposite and equal in magnitude, and the derivative of the total force is negative. The Taylor expansion in Eq \eqref{eq:TaylorExpansionSaddle} cannot be used in this analysis, since it gives a field that does not go to zero at $z = \pm\infty$. However, since we are mainly interested in the intensity of the beam along the $z$-axis, for $x' = y' = 0$, Eq. \eqref{eq:saddleIntensityRotFrame} yields the simple result
\begin{equation}
    \vert E_{s}\vert^2(z) = \frac{32P}{17\pi c\epsilon_0}\frac{1}{w_0^2(1+z^2/z_R^2)}.
\end{equation}
\noindent Thus, the gradient force along the $z$-axis is
\begin{equation}
    F_g(z) = -\frac{16P\Re\{\alpha\}}{17\pi c\epsilon_0}\frac{z}{ w_0^2 z_R^2(1+z^2/z_R^2)^2},
\end{equation}
\noindent For a silica particle, the extinction cross-section is well approximated by \cite{jones2015optical},
\begin{equation}
    \sigma_{\rm{ext}} \approx \frac{8\pi^3\vert\alpha\vert^2}{3\epsilon_0^2\lambda_0^4}.
\end{equation}
\noindent Therefore, the scattering force is
\begin{equation}
    F_s(z) =  \frac{128P\pi^2\vert\alpha\vert^2}{51\lambda_0^4\epsilon_0^2c}\frac{1}{w_0^2(1+z^2/z_R^2)}.
\end{equation}
Making $F_g(z) + F_s(z) = 0$ leads to a quadratic equation:
\begin{equation}
    z^2 - \frac{3\lambda_0^4\epsilon_0\Re\{\alpha\}}{8\pi^3\vert\alpha\vert^2}z + z_R^2 = 0.
\end{equation}
\noindent To have real solutions, the following inequality must be satisfied:
\begin{equation}
    \left(\frac{3\lambda_0^4\epsilon_0\Re\{\alpha\}}{8\pi^3\vert\alpha\vert^2}\right)^2 - 4z_R^2 > 0,
\end{equation}
\noindent where we disregarded the equality since it generates an unstable equilibrium point. Finally, by considering the particle's radius fixed, which fixes the polarizability, and using the definition of the Rayleigh range, we find a condition for $w_0$ to have stability along the $z$-direction:
\begin{equation}\label{eq:ScatteringStabilityCondition}
    w_0 < \sqrt{\frac{3\lambda_0^5\epsilon_0\Re\{\alpha\}}{16\pi^4\vert\alpha\vert^2}}.
\end{equation}
\noindent If the above condition is satisfied, there is a stable equilibrium position for the particle's longitudinal motion at
\begin{equation}\label{eq:z_eq}
    z_{\rm{eq}}= \frac{3\lambda_0^4\epsilon_0\Re\{\alpha\}}{16\pi^3\vert\alpha\vert^2} - \frac{1}{2}\sqrt{\left(\frac{3\lambda_0^4\epsilon_0\Re\{\alpha\}}{\pi^3\vert\alpha\vert^2}\right)^2 - 4z_R^2}.
\end{equation}

\section{Overdamped stability analysis}\label{sec:OverdampedAppendix}


In the overdamped regime, the Eqs. of motion simplify to,
\begin{subequations}\label{eq:OverdampedEq}
    \begin{align}
        0 &= (\Omega^2+\omega_x^2)x' - \gamma(\Dot{x}'-\Omega y') + 2\Omega\Dot{y}' + \frac{F_{\rm{th}}}{m},\label{eq:OverdampedX}\\
        0 &= (\Omega^2-\omega_y^2)y' - \gamma(\Dot{y}'+\Omega x') - 2\Omega\Dot{x}' + \frac{F_{\rm{th}}}{m} ,\label{eq:OverdampedY}\\
        0 &= -\omega_z^2 z - \gamma \Dot{z} +  \frac{F_{\rm{th}}}{m} \label{eq:OverdampedZ}.
    \end{align}
\end{subequations}

Similarly to Section \ref{sec:dampedStability}, let $x' = x'_0e^{\lambda t}$ and $y' = y'_0e^{\lambda t}$. By substitution in equations \eqref{eq:OverdampedX} and \eqref{eq:OverdampedY}, and neglecting the stochastic force term, we have
\begin{subequations}
    \begin{align}
    &(\Omega^2+\omega_x^2-\gamma\lambda)x'_0 + (\gamma\Omega+2\lambda\Omega)y'_0 = 0, \\
    &(\Omega^2-\omega_y^2-\gamma\lambda)y'_0 - (\gamma\Omega+2\lambda\Omega)x'_0 = 0.
\end{align}
\end{subequations}
\noindent Casting the equation in matrix form
\begin{equation}
    \begin{bmatrix}
        \Omega^2+\omega_x^2-\gamma\lambda & \gamma\Omega+2\lambda\Omega\\
        -\gamma\Omega-2\Omega\lambda & \Omega^2-\omega_y^2-\gamma\lambda
\end{bmatrix}
\begin{bmatrix}
        x'_0\\
        y'_0
\end{bmatrix} = \begin{bmatrix}
        0\\
        0
\end{bmatrix},
\end{equation}
\noindent and demanding that the matrix's determinant is equal to zero, one arrives at a quadratic polynomial for $\lambda$:
\begin{equation}
    a\lambda^2 + b\lambda + c = 0,
\end{equation}
\noindent where
\begin{subequations}
    \begin{align}
    a &=  \gamma^2+4 \Omega ^2,\\
    b & = 2 \gamma\Omega^2+\gamma\omega_y^2-\gamma\omega_x^2,\\
    c& =  \Omega ^4 +  \Omega ^2(\gamma ^2+\omega_x^2-\omega_y^2) -\omega_x^2 \omega_y^2.
\end{align}
\end{subequations}

Since the roots are $\lambda^{\pm} = -b/2a \pm \sqrt{b^2-4ac}/2a$ and $a,b>0$ (see Eq. \eqref{eq:omegaDefinition}), then they are complex with a negative real part if $b^2-4ac < 0$. If $b^2-4ac > 0$, one needs to require $-b + \sqrt{b^2-4ac} < 0$ so both roots are negative. Taking the worst case, assuming that $b^2-4ac > 0$ in conjunction with $a,b>0$, leads to requiring that $c > 0$, i.e.,
\begin{equation}\label{eq:quarticOverdamped}
    \Omega ^4 +  \Omega ^2(\gamma ^2+\omega_x^2-\omega_y^2) -\omega_x^2 \omega_y^2 > 0.
\end{equation}
\noindent This condition is interesting because it will force both roots to be complex, ending up in the first case.

Inequality \eqref{eq:quarticOverdamped} leads to demanding $\vert\Omega\vert > \omega^+$ for stability, where $\omega^+$ is the real positive root of the polynomial. Moreover, note that the polynomial necessarily has two real and two imaginary roots. Let $\Omega = \pm\sqrt{u}$, then the roots for the quadratic equation in $u$ are
\begin{equation}
    u^\pm = \frac{-\gamma^2+\omega_y^2-\omega_x^2\pm\sqrt{(\gamma ^2+\omega_x^2-\omega_y^2)^2+4\omega_x^2 \omega_y^2}}{2}.
\end{equation}
\noindent Since the argument inside the square root is positive, $u^+$ must give the real roots of Eq. \eqref{eq:quarticOverdamped}. Therefore, the stability criterion for the overdamped case
becomes identical to Eq. \eqref{eq:stability_criteria}.
\begin{figure*}[htb!]
    \includegraphics{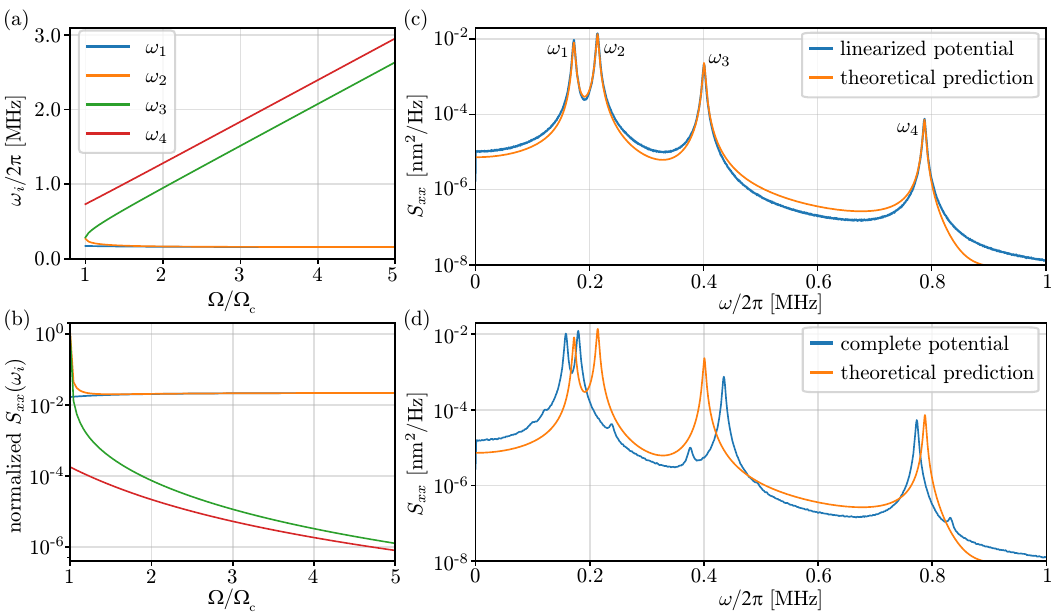}
    \caption{a) Relation between the natural frequencies and b) power, normalized by $\rm{max}\lbrace S_{xx}(\omega_1)\rbrace$, to the saddle's angular rotation speed $\Omega$. The color schemes employed in panels a) and b) are the same. The angular frequencies $\omega_i$ follow the order from lower to higher as established in panel c).  c) and d) show examples of simulated position PSDs for the linearized and complete optical potential in the laboratory reference frames.}
    \label{fig:PSD_Panel}
\end{figure*}
\section{Numerical simulation parameters}\label{sec:simulationsDetails}

For all the simulations described in Section \ref{sec:simulations}, the particle is initially located at
\begin{equation}\label{eq:simInitialPosition}
    \mathbf{r}_0 = \begin{bmatrix}
        \sqrt{k_bT/m\omega_{x,h}^2}\\
        \sqrt{k_bT/m\omega_{y,h}^2}\\
        z_{eq}+\sqrt{k_bT/m\omega_{z,h}^2}
\end{bmatrix},
\end{equation}
\noindent where $\sqrt{k_bT/m\omega_{i,h}^2}$ is the standard deviation of the particle's position i when it is trapped by a standard Gaussian beam optical tweezer (OT) (considering a harmonic approximation). The $\omega_{i,h}$ are the natural frequencies for a harmonic trap given by
\begin{equation}
    \begin{bmatrix}
        \omega_{x,h}^2 \\
        \omega_{y,h}^2 \\
        \omega_{z,h}^2
    \end{bmatrix}
    =
    \frac{2P\Re\{\alpha\}}{c\epsilon_0\pi w_0^2m}
    \begin{bmatrix}
        2/w_0^2 \\
        2/w_0^2 \\
        1/z_R^2
    \end{bmatrix}.
\end{equation}
\noindent The $z_{eq}$ term is the equilibrium position along the optical axis, which is different from the zero due to the scattering force. Most interestingly, its value for a Gaussian beam and the proposed saddle beam is the same, given by Eq. \eqref{eq:z_eq} derived in Appendix \ref{sec:ScatterAppendix}. This arises because, at the optical axis, only the Gaussian $E^{LG}_{0,0}$ term has a non-vanishing value for the saddle beam. 
The initial velocity is given by the standard deviation of velocity for a particle in a harmonic trap, given by $v_0 = \sqrt{k_BT/m}$ for all directions. The main motivation behind this choice of initial conditions is the idea that one could first trap a particle using a standard Gaussian beam, and later switching to the saddle beam.

For the beam's overall parameters, we set its wavelength at $\SI{1550}{nm}$, the numerical aperture (NA) at $0.6$, and the optical power at $\SI{500}{mW}$. The low NA is required to prevent the structure of the beam's intensity from collapsing to a non-paraxial Gaussian-like shape. The effect of high focusing  was numerically evaluating the transmission of the saddle beam by a high-NA lenses (NA $>0.8$) using the angular spectral representation formalism \cite{novotny2012principles}, and we have observed that an NA$ = 0.6$ is a good compromise between focusing and keeping the structure of the saddle superposition. We considered SiO$_2$ nanoparticles with a $\SI{150}{nm}$ radius. The gas pressure was set to $\SI{10}{mbar}$. The scattering force given by the second term of Eq. \eqref{eq:opticalForce} was included in the simulations. 

All simulations were performed using the Runge-Kutta 4$^{\mathrm{th}}$ order algorithm with a simulation step of $\SI{10}{ns}$ and 1,000,000 steps.

\section{Position power spectral density}\label{sec:PSD}

We now evaluate the particle's position PSD in the rotating reference. We will focus on the transverse dynamics given by the equations \eqref{eq:dampedXeq}, and \eqref{eq:dampedYeq} with $F_i = F_{th,i}$ obeying equations \eqref{eq:thermalCorrelation}, since the PSD for the longitudinal motion is well-known:
\begin{equation}\label{eq:zPSD}
S_{zz}(\Omega)= \frac{2\gamma_m k_B T}{m[(\Omega^2-\omega_z^2)^2+\gamma^2\omega_z^2]}.
\end{equation}
\noindent We start by expressing the differential equations in the frequency domain,
\begin{subequations}
\begin{align}
    A_x\Tilde{x}'(\omega) &= B\Tilde{y}'(\omega) + \Tilde{F}_{th}(\omega)/m, \\
    A_y\Tilde{y}'(\omega) &= -B\Tilde{x}'(\omega)+ \Tilde{F}_{th}(\omega)/m,
\end{align}
\end{subequations}
\noindent where the tilde symbolizes the variables in the frequency domain and we have,
\begin{subequations}
\begin{align}
    A_x &= -\Omega^2-\omega_x^2-\omega^2-i\omega\gamma, \\
    A_y &= -\Omega^2+\omega_y^2-\omega^2-i\omega\gamma, \\
    B &=\gamma\Omega-i2\omega\Omega,
\end{align}
\end{subequations}
\noindent We consider the stochastic thermal force has the same frequency spectrum in both directions of the transverse plane. Since the above equations are linear in $\Tilde{x}'$ and $\Tilde{y}'$, one can solve explicitly to find
\begin{subequations}\label{eq:fourierTransformRotatedFrame}
\begin{align}
    \Tilde{x}'(\omega) &= \chi_x(\omega)\Tilde{F}_{th} \\
    \Tilde{y}'(\omega) &=  \chi_y(\omega)\Tilde{F}_{th},
\end{align}    
\end{subequations}
\noindent where $\chi_i(\omega)$ are the mechanical susceptibilities in the rotating reference frame, given by
\begin{subequations}
    \begin{align}
    \chi_x(\omega) &= \frac{1}{m}\frac{A_y+B}{B^2+A_xA_y},\\
    \chi_y(\omega) &= \frac{1}{m}\frac{A_x-B}{B^2+A_xA_y}.
\end{align}
\end{subequations}
\noindent Finally, the position PSDs in the rotating reference frame are
\begin{subequations}
\begin{align}
    S_{x'x'}(\omega) &= 2m\gamma k_BT\vert\chi_x(\omega)\vert^2, \\
     S_{y'y'}(\omega) &= 2m\gamma k_BT\vert\chi_y(\omega)\vert^2,
\end{align}
\end{subequations}
\noindent where we used
\begin{subequations}
    \begin{align}
        &S_{ii}(\omega) = \vert\Tilde{i}(\omega)\vert^2, \\
        &\vert\Tilde{F}_{th}\vert^2 = S_{\Tilde{F}_{th}\Tilde{F}_{th}}(\omega) = 2m\gamma k_BT.
    \end{align}
\end{subequations}

One can use the Fourier transform of the positions in the rotating reference frame to evaluate the form of the PSDs in the laboratory reference frame. We can show that,
\begin{subequations}
    \begin{align}
    F.T.\{f(t)\cos(\Omega t)\} &= \frac{1}{2}(\Tilde{f}(\omega^+) + \Tilde{f}(\omega^-)), \\
    F.T.\{f(t)\sin(\Omega t)\} &= \frac{1}{2i}(\Tilde{f}(\omega^+) - \Tilde{f}(\omega^-)),
\end{align}
\end{subequations}
\noindent where $F.T.\{\cdot\}$ is the Fourier transform and $\omega^\pm = \omega\pm\Omega$. Using this result with the coordinate transformation 
\begin{align}\label{eq:cordinateTransformation}
    x'&=x\cos\theta+y\sin\theta, \\
    y'&=y\cos\theta-x\sin\theta,
\end{align}
\noindent one can relate the Fourier transform of the positions from one coordinate to the other,
\begin{subequations}
    \begin{align}
    \begin{split}
        \Tilde{x}(\omega) = \frac{1}{2}(\Tilde{x}'(\omega^+)&+\Tilde{x}'(\omega^-)) \\&- \frac{1}{2i}(\Tilde{y}'(\omega^+)-\Tilde{y}'(\omega^-)),
        \end{split}\\
    \begin{split}
        \Tilde{y}(\omega) = \frac{1}{2i}(\Tilde{x}'(\omega^+)&-\Tilde{x}'(\omega^-)) \\&+ \frac{1}{2}(\Tilde{y}'(\omega^+)+\Tilde{y}'(\omega^-)).
    \end{split}
\end{align}
\end{subequations}
\noindent Note that $\Tilde{F}_{th}(\omega^\pm) = \Tilde{F}_{th}(\omega) = \rm{costant}$ since the thermal force is a white noise. Thus, using Eq. \eqref{eq:fourierTransformRotatedFrame},
\begin{subequations}
    \begin{align}
    \begin{split}
        \Tilde{x}(\omega) = \frac{\Tilde{F}_{th}}{2}\Big(&\chi_x(\omega^+)+\chi_x(\omega^-) \\+&i(\chi_y(\omega^+)-\chi_y(\omega^-))\Big),
    \end{split}\\
    \begin{split}
        \Tilde{y}(\omega) = \frac{\Tilde{F}_{th}}{2}\Big(&\chi_y(\omega^+)+\chi_y(\omega^-)\\ + &i(\chi_x(\omega^-)-\chi_x(\omega^+))\Big).
    \end{split}
\end{align}
\end{subequations}
\noindent Finally, the particle's position PSD in the laboratory reference frame is given by
\begin{subequations}\label{eq:PSD_labframe}
    \begin{align}
    \begin{split}
        S_{xx} = \frac{m\gamma k_BT}{2}\Big\vert\Big(&\chi_x(\omega^+)+\chi_x(\omega^-)\\ +&i(\chi_y(\omega^+)-\chi_y(\omega^-))\Big)\Big\vert^2,
    \end{split}\\
    \begin{split}
        S_{yy} = \frac{m\gamma k_BT}{2}\Big\vert\Big(&\chi_y(\omega^+)+\chi_y(\omega^-) \\+ &i(\chi_x(\omega^-)-\chi_x(\omega^+))\Big)\Big\vert^2.
    \end{split}
    \end{align}
\end{subequations}

One could find the frequencies that maximize the position's PDF to find the natural frequencies of the system's dynamics. However, it is easier to solve the eigenvalue problem of the system of differential equations given by \eqref{eq:dampedXeq} and \eqref{eq:dampedYeq} considering zero damping. The eigenvalues will provide the natural frequencies in the rotating frame. According to equations \eqref{eq:PSD_labframe}, the frequencies in the laboratory frame are shifted by $\pm\Omega$. To achieve this, we express the system as the matrix equation

\begin{equation}
	\frac{d}{dt}\begin{bmatrix}
		x' \\
		y' \\
		\dot{x}' \\
		\dot{y}' \\
	\end{bmatrix}
	=
\begin{bmatrix}
	0 & 0 & 1 & 0 \\
	0 & 0 & 0 & 1 \\
	\Omega^2+\omega_x^2 & 0 & 0 & 2\Omega \\
	0 & \Omega^2-\omega_y^2 & -2\Omega & 0
\end{bmatrix}
\begin{bmatrix}
		x' \\
		y' \\
		\dot{x}' \\
		\dot{y}' \\
	\end{bmatrix},
\end{equation}
\noindent There are two conjugate pairs of eigenvalues for the dynamics matrix given by
\begin{subequations}
\begin{align}
    \lambda_{1,\pm} &= \pm i\frac{\sqrt{\vert C-\sqrt{D}\vert}}{\sqrt{2}}, \\
     \lambda_{2,\pm} &= \pm i\frac{\sqrt{\vert C+\sqrt{D}\vert}}{\sqrt{2}},
\end{align}
\end{subequations}
\noindent where
\begin{subequations}
\begin{align}
    C &= \omega_x^2-\omega_y^2-2\Omega^2,\\
    D &= w_x^4+w_y^4 +2w_x^4w_y^4 + 8\Omega^2(w_y^2-w_x^2).
\end{align}
\end{subequations}
Finally, considering the positive-valued frequencies in the laboratory frame, they are given by
\begin{subequations}
\begin{align}
    \omega_1 &= \lambda_{1,+}/i - \Omega, \\
    \omega_2 &= \lambda_{2,-}/i + \Omega, \\
    \omega_3 &= \lambda_{2,+}/i + \Omega, \\
    \omega_4 &= \lambda_{1,+}/i + \Omega.
\end{align}
\end{subequations}

Figure \ref{fig:PSD_Panel}a) illustrates how the frequencies $\omega_i$ change with the speed of the saddle's rotation, while Figure \ref{fig:PSD_Panel}b) displays the power for each frequency normalized by $\rm{max}\lbrace S_{xx}(\omega_1)\rbrace$. Considerations for the $y$-direction are the same, thanks to symmetry. We can categorize the natural frequencies into two sets: slow frequencies $\lbrace\omega_1,\omega_2\rbrace$ and fast frequencies $\lbrace\omega_3,\omega_4\rbrace$. As $\Omega$ increases, the slow frequencies approach a common value and power. Conversely, the fast frequencies increase, and their power asymptotically approaches zero. Figures \ref{fig:PSD_Panel}c) and \ref{fig:PSD_Panel}d) depict examples of simulated position PSDs for the linearized and complete optical potential in the laboratory reference frames for the parameters contained in Table \ref{tab:simulationParameters} and $\Omega/\Omega_c = 1.1$. The blue curves stem from an ensemble of 200 averages of individual PSDs, while the orange curves represent the theoretical prediction from Eq. \eqref{eq:PSD_labframe}. The linearized model plot demonstrates a strong agreement between theory and simulation. The disagreements may arise due to the finite observation time considered in the simulations. However, due to its nonlinear nature, the complete optical potential exhibits shifted natural frequencies and additional peaks \cite{suassuna2021path}.

\end{document}